\documentclass[%
reprint,
amsmath,amssymb,
aps,
]{revtex4-2}

\usepackage{graphicx}% Include figure files
\usepackage{subfigure}
\usepackage{dcolumn}% Align table columns on decimal point
\usepackage{bm}% bold math
\usepackage{upgreek}
\begin{document}

\preprint{APS/123-QED} 
%\preprint{}

\title{Sustainable early-stage lasing in a low-emittance electron storage ring}

\author{Kaishang Zhou}
 \email{zhoukaishang@whu.edu.cn}
\altaffiliation[Current address ]{The Institute for Advanced Studies, Wuhan University}
%\affiliation{The Institute for Advanced Studies, Wuhan University, Wuhan 430072 , China}
\affiliation{Department of Engineering Physics, Tsinghua University, Beijing 100084 , China}
\author{Zhenghe Bai}
%\email{baizhe@ustc.edu.cn}
\affiliation{National Synchrotron Radiation Laboratory, USTC, Hefei 230029, China}
\author{Renkai Li}
\email{lirk@tsinghua.edu.cn}
\affiliation{Department of Engineering Physics, Tsinghua University, Beijing 100084 , China}

\date{\today}

%\tableofcontents

\begin{abstract}
In this Letter, we report on the concept and analysis of a low-emittance electron storage ring, in which the electron beams undergo an early-stage self-amplified spontaneous emission lasing process on a turn-by-turn basis. The lasing process for each pass through a long undulator in the ring is terminated when the radiated power is still negligible compared to the total synchrotron loss of each circulation, and the electron beams can be maintained in an equilibrium state that supports sustainable lasing. A self-consistent model is derived for evaluation of the properties of the electron beams, and a design with numerical modeling is presented that demonstrates the feasibility of generating short-wavelength radiation at the kW power level.

\end{abstract}

%\keywords{}

\maketitle

%\section{\label{sec:level1}Introduction}
Electron beam-based short wavelength radiation sources, including storage ring light sources and free-electron lasers, have emerged as transformative research tools for scientific discoveries and technological innovations \cite{wille91,madey71,emma09}. The underlying physics, i.e., the mechanism of extracting energy from the electron beam into electromagnetic radiation with desired properties, is itself a deep and forefront research field. In an electron storage ring, electron beams generate spontaneous radiation with low peak power at a high repetition rate. In a linac-based high gain free-electron lasers (FEL), the interaction between high-brightness electron beams and the radiation field leads to extremely high peak power at a relatively low repetition rate. Aiming at combining the high repetition rate and high peak power, the pursuit of alternate schemes, including energy-recovery linacs, superconducting linacs, and storage-ring FELs (SRFEL), has accumulated considerable advances in recent decades.

The concept of a storage-ring FEL, in which electron beams circulate with a long lifetime and maintain adequate qualities for lasing in each pass, was proposed and successfully demonstrated at a long wavelength over the infrared to ultraviolet (UV) range in the low-gain regime using optical cavities \cite{Couprie97, Neil2003}. The radiation wavelength of the SRFEL toward shorter wavelengths was hindered by the lack of suitable oscillator mirrors \cite{litvinenko2001, Trovo2002, Wu06}, and the radiation power is essentially restricted by the small signal gain in each pass \cite{Renieri79, Renieri80}.

Aiming at reaching both shorter wavelength and higher average power \cite{wangner10}, the high-gain SRFEL scheme was proposed \cite{Kim85, Murphy1985, Cornacchia86, N92}, which utilizes a dedicated bypass line to decouple the FEL interaction with the main storage ring. Each stored electron beam will be switched into the bypass line for lasing amplification and then guided back into the main storage ring for damping until its quality is adequate for another lasing pass. The main advantage of this scheme is that it can produce short wavelength radiation with high peak power. However, the average radiation power is still low due to the effective repetition rate of lasing being limited by the necessary damping process. Meanwhile, the beam energy spread in a high energy storage ring is too large for effective FEL lasing, and thus, a transverse gradient undulator (TGU) was proposed to tackle this challenge \cite{Cai13,huang12}.

Another type of high-gain SRFEL without a bypass line was investigated in which the beam dynamics of high-gain lasing and the storage ring itself are coupled \cite{huang08, Dattoli12}. The coupled beam dynamics lead to an equilibrium state, and the average spectral brightness can potentially be increased by 2 to 3 orders of magnitude over spontaneous radiation. This scheme operates with a long undulator ($\sim $100 m) and a high bunch peak current ($ \sim $300 A). The average current is limited to 10 mA resulting from a low multi-bunch filling pattern \cite{Mitri15, Cai13,sy01}. The lasing performance can be significantly improved with a seed laser, however, a suitable seed laser with a short wavelength and high repetition rate is technically challenging and not yet available \cite{Lee20}.

Since a high peak current is desirable for lasing, an RF longitudinal focusing insertion in a bypass line was introduced to increase the beam peak current to 450$\sim$700 A using an RF-induced energy chirp within the electron beam \cite{Agapov14, Agapov15}. The electron beam will be compressed to a high peak current before entering the undulator for lasing and then decompressed to circulate in the ring. This scheme requires a long RF linac with a large energy gain, which also sets the repetition rate of the lasing pass operation or requires superconducting linacs at high cost \cite{Mitri15, Mitri21}.

In this Letter, we discuss the concept and analysis of an alternate SRFEL scheme in which electron beams undergo a sustainable early-stage self-amplified spontaneous emission (SASE) lasing process~\cite{saldin1980, Bonifacio, kroll78} each and every turn in a low-emittance storage ring. Although the peak radiation power is moderate in this new working regime, the ultimate repetition rate still leads to high average power at short wavelengths. There are a few important features of the proposed scheme:

First, the beam emittance of the storage ring is extremely low, which supports efficient SASE lasing for the given radiation wavelength ($4 \pi \varepsilon_{x} \leqslant\lambda_{s}$). Therefore, the beam peak current can be relaxed to $\sim$100 A, which is routinely achievable in modern storage rings, instead of the multi-hundred amperes required in previous studies to maintain a manageable gain length.  

Next, in addition to the emittance and beam current, energy spread is another key parameter to enable efficient SASE lasing. The equilibrium energy spread taking into account the coupled lasing process and storage ring dynamics, as we will discuss in detail later, is also optimized to a small value for our proposed scheme, and consequently, a TGU or a seed laser is no longer necessary.

Importantly, to enable sustainable, turn-by-turn lasing, we deliberately terminate the lasing process in each pass at an early stage so that electron beam qualities are essentially preserved through the undulator. More specifically, we control the quantum excitation caused by FEL lasing in each turn to be at least one order of magnitude lower than that of synchrotron radiation in the storage ring. The beam properties will be dominated by the ring dynamics and eventually reach equilibrium values close to those when no lasing occurs.

%\section{Layout and analytical model}

\begin{figure}[h]
	\includegraphics[scale=0.41]{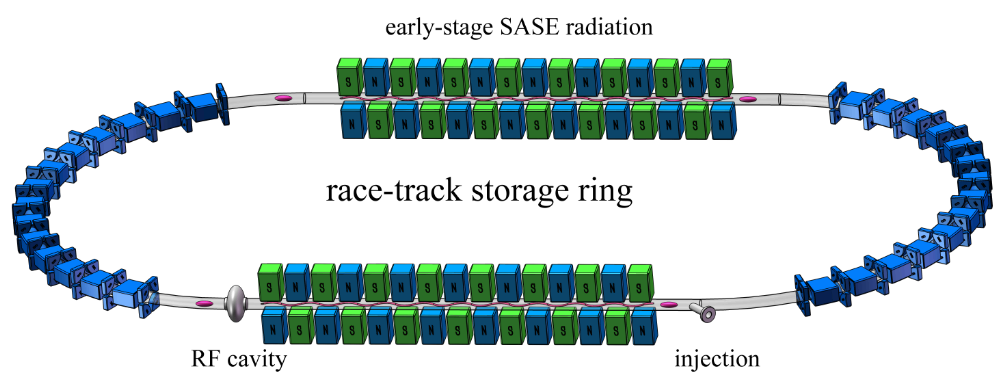}
	\caption{\label{fig:imag1} The electron beam circulates around a race-track storage ring that is supplied by an external injection, sustainable early-stage lasing happens in a long straight section, and the RF cavity compensates for the energy loss due to synchrotron radiation and early-stage SASE.}
\end{figure}

The layout of the proposed SRFEL scheme is shown in Fig.\;\ref{fig:imag1}. We have adopted a race-track configuration with two periods and two long straight sections, and each of the long straight sections hosts a long undulator with beta-matching/focusing for early-stage lasing.\\
%and Fig.~\ref{fig:epsart}.%
\\
The early-stage SASE is treated as a perturbation to the storage ring dynamics. The quantum excitation effect of the early-stage SASE in each turn will lead to an increase in energy spread. Then, the evolution of the energy spread in the storage ring can be written as \cite{Wo14, pk18}   
 \begin{equation}
 	\frac{d \sigma_{\delta}^{2}}{d t}=\frac{2}{\tau_{s}}\left(\sigma_{\delta0}^{ 2}-\sigma_{\delta}^{2}\right)+\frac{2}{T_{fel}} \sigma_{\delta}^{2}\label{new1}
 \end{equation}
where $\tau_{\mathrm{s}}$ is the natural damping time for the longitudinal motion of the proposed storage ring, while the damping effect of the early-stage SASE, as will be shown later, is negligible because the synchrotron radiation from bending magnets and long undulators is dominant in the storage ring. $\sigma_{\delta0}$ is the natural equilibrium energy spread, $\sigma_{\delta}$ is the new energy spread after considering the quantum excitation effect of the early-stage SASE, and ${T_{fel}}$ is characteristic growth time of the energy spread caused by the early-stage SASE.

When reaching the steady state condition $d\sigma_{\delta}^{2}/dt=0$, the new equilibrium energy spread can be expressed as
 \begin{equation}
 	\sigma_{\delta}^{2}=\frac{\sigma_{\delta0}^{2}}{1-\tau_{s} / T_{fel}}\label{new2}
 \end{equation}
 The natural equilibrium energy spread of a storage ring is given below \cite{Liu08,sands70, Wiedemann2015, Chao20}   
\begin{equation}
	\sigma_{\delta0}^{ 2}=\frac{1}{4 E_{0}^{2}} \tau_{\mathrm{s}} Q_{s}\label{key18}
\end{equation}
where, $ E_{0} $ is the electron beam energy of the storage ring, and $ Q_{s} $ is the parameter representing quantum excitation caused by synchrotron radiation, which is defined as $Q_{s}=\int_{0}^{\infty} n(\varepsilon) \varepsilon^{2} d \varepsilon$. Here, $n\left(\varepsilon\right)$ is the probability of emitting photons of energy $ \varepsilon $ per unit time.

$1/{T_{fel}}$ is defined as
\begin{equation}
	\frac{1}{{T_{fel}}}=\frac{1}{2\sigma_{\delta}^{2}}	\frac{d \sigma_{\delta}^{2}}{d t}\label{new4}
\end{equation}
Then, we introduce a parameter $ Q_{f} $ to quantify the quantum excitation caused by the early-stage SASE, which is defined as $ Q_{f}=\int_{\varepsilon_{f-\Delta}}^{\varepsilon_{f+\Delta}} n\left(\varepsilon_{f}\right) \varepsilon_{f}^{2} d \varepsilon_{f}$. Here, $ n\left(\varepsilon_{f}\right) $ is the probability of emitting photons of energy $ \varepsilon_{f} $ per unit time.  

Thus, the growth rate of the energy spread due to early-stage SASE excitation in the storage ring can be expressed as
\begin{equation}
	\frac{d \sigma_{\delta}^{2}}{d t} = \frac{Q_{f}}{2E_{0}^{2}}\label{high6}
\end{equation}
 Inserting Eq.(\ref{key18}) and Eq.(\ref{high6}) into Eq.(\ref{new2}), we obtain the ratio between the new equilibrium energy spread due to coupled dynamics and the natural equilibrium energy spread with the ring dynamics alone 
 \begin{equation}
	\frac{\sigma_{\delta}^{2}}{\sigma_{\delta0}^{2}}=\frac{1}{1-Q_{f} / Q_{s}}\label{new5}
\end{equation}
To control the lasing-induced growth of the equilibrium energy spread $\sigma_{\delta}$ over the original value, $Q_{f} / Q_{s} $ must be a small quantity. In the first-order approximation, the new equilibrium energy spread $\sigma_{\delta}$ can be expressed as        
\begin{equation}
	\frac{\sigma_{\delta}}{\sigma_{\delta0}}=\sqrt{1+\frac{Q_{f}}{Q_{s}}}\label{new3}
\end{equation}

In our proposed scheme, the synchrotron radiation power losses consist of two parts: the bending magnets and the long undulators. Hence, the quantum excitation $ Q_{s} $ can be expressed as $ Q_{s}= Q_{b}+Q_{w} $, where $ Q_{b} $ is from bending magnets and $ Q_{w} $ is from long undulators. Thus, Eq.(\ref{new3}) can be re-written as

\begin{equation}
	\frac{\sigma_{\delta}}{\sigma_{\delta0}}=\sqrt{1+\frac{Q_{f}}{Q_{b}+Q_{w}}}\label{now1}
\end{equation}

It is well-established in synchrotron radiation theory that $Q_{b}\approx 1.36U_{b} \varepsilon_{c}$, where $ U_{b} $ is the average energy loss radiated by bending magnets per electron per turn, and $\varepsilon_{c}  $ is the critical photon energy of bending magnets. Similarly, the synchrotron radiation from long undulators can be treated as a series of small strong bending magnets, and thus $Q_{w}\approx 1.36U_{w} \varepsilon_{w}$, where $ U_{w} $ is the average energy loss radiated in the undulators per electron per turn, and $\varepsilon_{w}  $ is the equivalent critical photon energy of undulators.
 
To quantify the quantum excitation induced by the early-stage SASE, we make the following assumptions: take the radiated photon energy $ \varepsilon_{f} $ as a constant value since the bandwidth of FEL lasing is narrow. Then, $N_{f}=\int_{\varepsilon_{f-\Delta}}^{\varepsilon_{f+\Delta}} n\left(\varepsilon_{f}\right) d \varepsilon_{f}$ is the total number of photons per electron per pass. Therefore, $ Q_{f}=N_{f} \varepsilon_{f}^{2}  $ and $ N_{f}= U_{f}/\varepsilon_{f}$. Thus, $Q_{f}\approx U_{f} \varepsilon_{f}$.  Here, $ U_{f} $ is the average energy loss per electron per pass. Combining the above formulas, we obtain the ratio in Eq.(\ref{now1}) as follows \\
\begin{equation}
	\frac{\sigma_{\delta}}{\sigma_{\delta0}}=\sqrt{1+\frac{U_{f}\varepsilon_{f}}{1.36(U_{b}\varepsilon_{c}+U_{w}\varepsilon_{w}) } }\label{key30}
\end{equation}

We aim at maintaining the new equilibrium energy spread sufficiently low to support sustainable turn-by-turn lasing at an appreciable power level. It is straightforward to conclude the two criteria from Eq.(\ref{key30}). First, the synchrotron radiation energy loss from bending magnets and undulators should dominate over the early-stage SASE, which means $ U_{b}, U_{w}\gg U_{f}$, and this criterion is self-consistent with the assumption in  Eq.(\ref{new1}). Second, the critical photon energy of the bending magnets and the equivalent critical photon energy of undulators should be much larger than the lasing photon energy, i.e. $\varepsilon_{c},\varepsilon_{w} \gg \varepsilon_{f}$. 

In addition to the criteria mentioned above for sustainable FEL lasing, we need to design and optimize a ring lattice for the proposed SRFEL scheme. The primary goal is to achieve desirable beam parameters that support SASE lasing in a reasonable undulator length. 

To design a low-emittance storage ring, we have taken the following strategy. First, multi-bend achromat (MBA) cells are used in the race-track arcs to achieve low emittance. It is well established that when a long undulator is inserted in a region with non-zero dispersion, the beam emittance will considerably grow, therefore, long undulators are inserted in dispersion-free straight sections. Such an arrangement has an additional benefit related to the increased radiation damping, which leads to reduced beam emittance in the ring.

\begin{figure}[h] 
	\includegraphics[width=8.5cm,height=4.5cm]{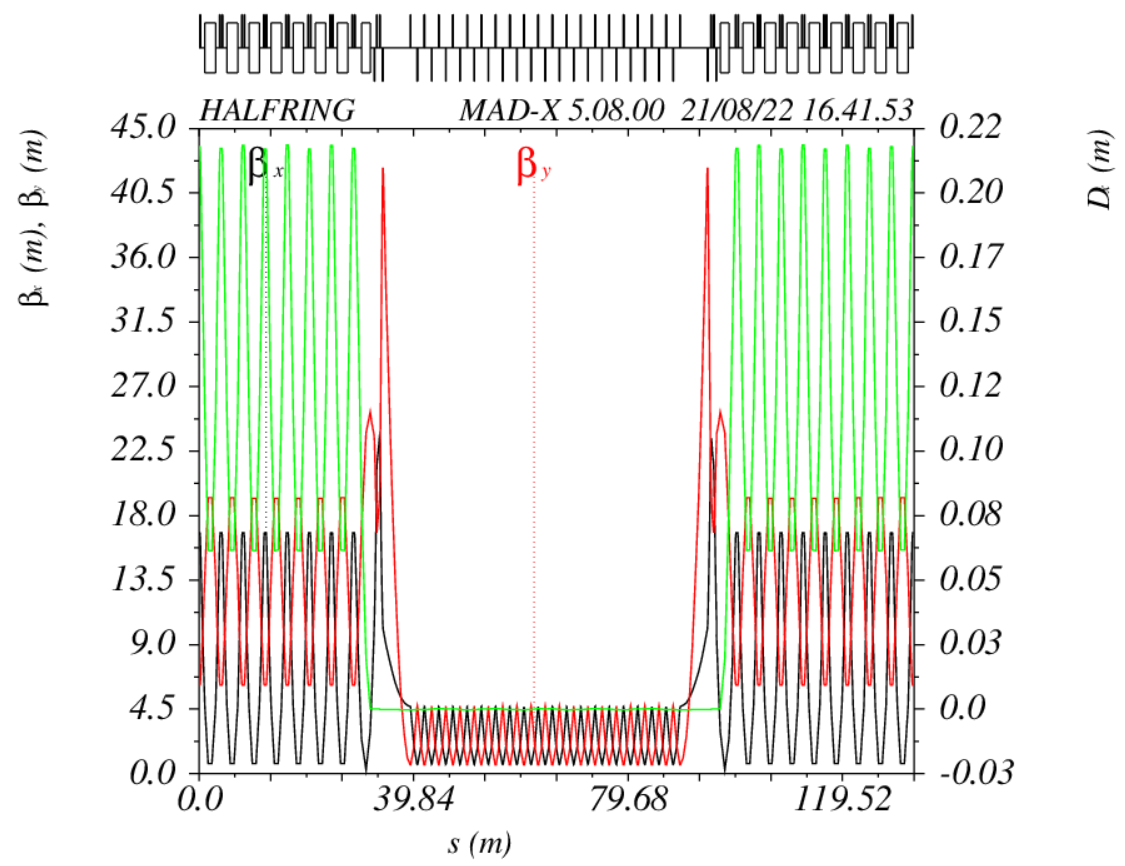}% Here is how to import EPS art
	\caption{\label{fig:imag3} The lattice of one period of the storage ring}
\end{figure}

For such a low-emittance storage ring, the effect of IBS, which might potentially blow up the natural energy spread and emittance, should be carefully considered. In the design of the storage ring, we adopted a commonly used full-coupling, i.e., round electron beam configuration, and used the well-developed and benchmarked Bj-SM model in the particle tracking code ELEGANT \cite{Borland} to evaluate the IBS effect. 

The beam energy is a key parameter in the design of a storage ring, which is largely determined by the required radiation wavelength range. As the beam energy increases, the energy spread decreases and the emittance grows with the IBS effects taken into account, even though high beam energy is effective in suppressing particle-particle interactions. The energy spread and emittance need to be balanced to obtain a suitable 3D gain length for the proposed early-stage SASE process.

\begin{table}[h]
	\caption{\label{tab:table3}%
		Parameters of the electron beam and undulator for numerical simulation in the storage ring }
	\begin{ruledtabular}
		\begin{tabular}{ccc}
			Parameter&Values&Unit
			\\
			%\mbox{Three}&\mbox{Four}&\mbox{Five}\\
			\hline
			Beam energy &1.6&\mbox{GeV}\\
			Circumference &265&\mbox{m} \\
			Energy loss per turn (with insertion)&349.2&\mbox{keV} \\
			Damping time ($\tau_{\mathrm{x}}$,$\tau_{\mathrm{y}}$,$\tau_{\mathrm{s}}$)&7.5/ 8.1/4.2&\mbox{ms} \\
			Momentum compaction factor&$1.7$&\mbox{$ 10^{-3} $}  \\
			Emittance (with IBS $\varepsilon_{x}=\varepsilon_{y})$&264&\mbox{pm-rad}  \\
			Energy spread (with IBS) &$7.3156$&\mbox{$ 10^{-4} $} \\
			Energy spread (new equilibrium) &$7.3209$&\mbox{$ 10^{-4} $} \\
			Bunch charge &2&\mbox{nC } \\
			Bunch length (RMS)&2.3&\mbox{mm}  \\
			Peak current &103&\mbox{A } \\
			Undulator period &3.0&\mbox{cm}\\
			Undulator parameter $ K $&3.95&\mbox{ } \\
			Radiation Wavelength &13.5&\mbox{nm}  \\
		\end{tabular}
	\end{ruledtabular}
\end{table} 

% \section{Design example of kW EUV light source}
Combining all the above considerations, and typical for designing and optimizing a system with multiple objects and multiple parameters of coupled or even competing effects, several iterations are used to obtain a reasonable solution.

In the following, we will present an example of the proposed sustainable early-stage SASE SRFEL scheme, including the parameters of the storage ring and modeling of the FEL process. Numerical results demonstrate the feasibility of such a scheme capable of generating kW-level power radiation at the EUV wavelength. 

The optimized ring design operates at 1.6 GeV beam energy with a circumstance of 265 m. Each of the long straight sections accommodates an undulator of 45.6 meters in length. The equilibrium beam parameters of the storage ring are calculated by ELEGANT. Fig.\;\ref{fig:imag3} shows the lattice of one period of the storage ring. The main parameters of the proposed scheme are summarized in Table \ref{tab:table3}. 
The emittance, energy spread, and bunch peak current, which are particularly relevant to lasing, are optimized to 264 pm-rad, $7.3156\times10^{-4}$ and 103 A, respectively.

\begin{figure}[h]
	\includegraphics[scale=0.35]{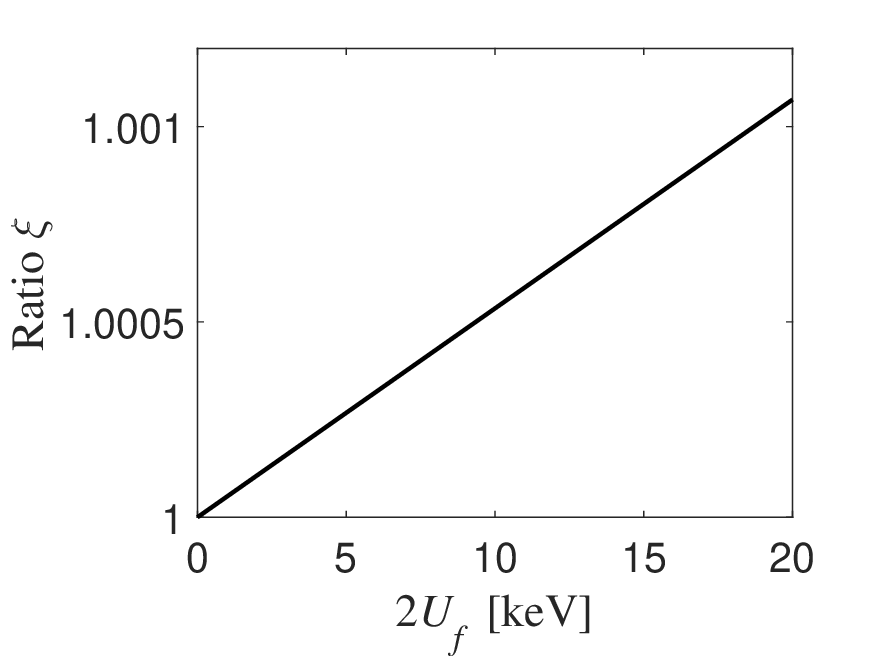}% Here is how to import EPS art
	\caption{\label{fig:imag2}  Ratio $ \xi $ changes over the one turn energy loss $ 2U_{f} $ caused by FEL lasing in the storage ring}
\end{figure}

The target lasing wavelength of our design is 13.5 nm, corresponding to a photon energy of $\varepsilon_{f}=0.092$ keV. The critical photon energy of the bending magnets is $\varepsilon_{c}=0.87 $ keV and the equivalent critical photon energy of undulators is  $\varepsilon_{w}=1.69 $ keV. The energy loss per electron per turn from the bending magnets is $U_{b}=55.6$ keV, and that from the undulators is $U_{w}=293.6$ keV. We assume that the energy loss due to lasing per electron per pass from each undulator is $ U_{f} $. Since the proposed storage ring hosts two periods and two undulators, the total lasing energy loss per electron per turn is $2 U_{f} $. We use Eq. \ref{key30} to evaluate the increased beam energy spread after taking into account the sustainable early-stage SASE process. The ratio $ \xi $ ($\sigma_{\delta}/\sigma_{IBS}$) between the new equilibrium energy spread and the original equilibrium energy spread with IBS of the storage ring itself, is plotted in Fig.\;\ref{fig:imag2} as a function of $2 U_{f} $.\\

\begin{figure}[h]
 	\includegraphics[scale=0.28]{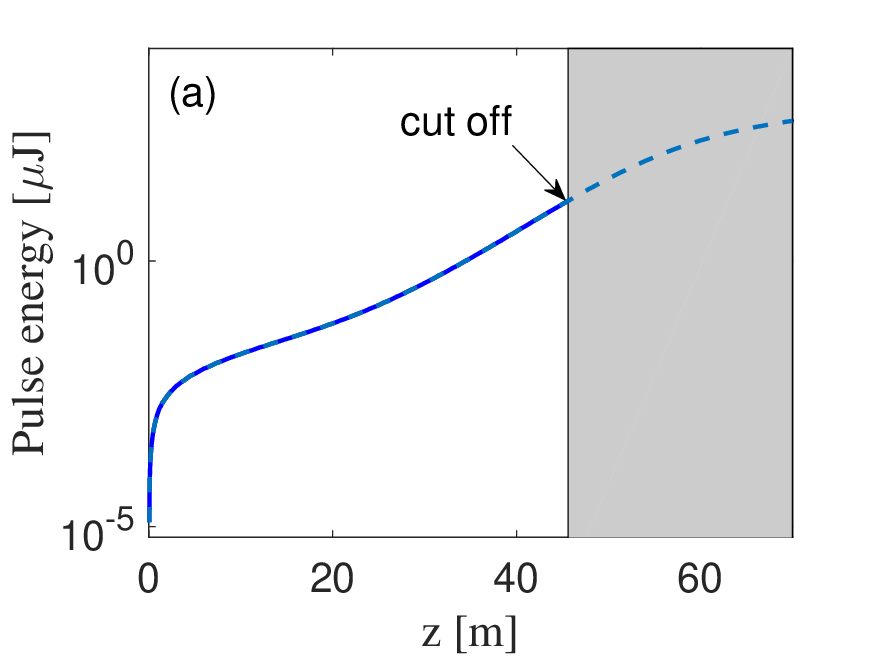}
 	\includegraphics[scale=0.28]{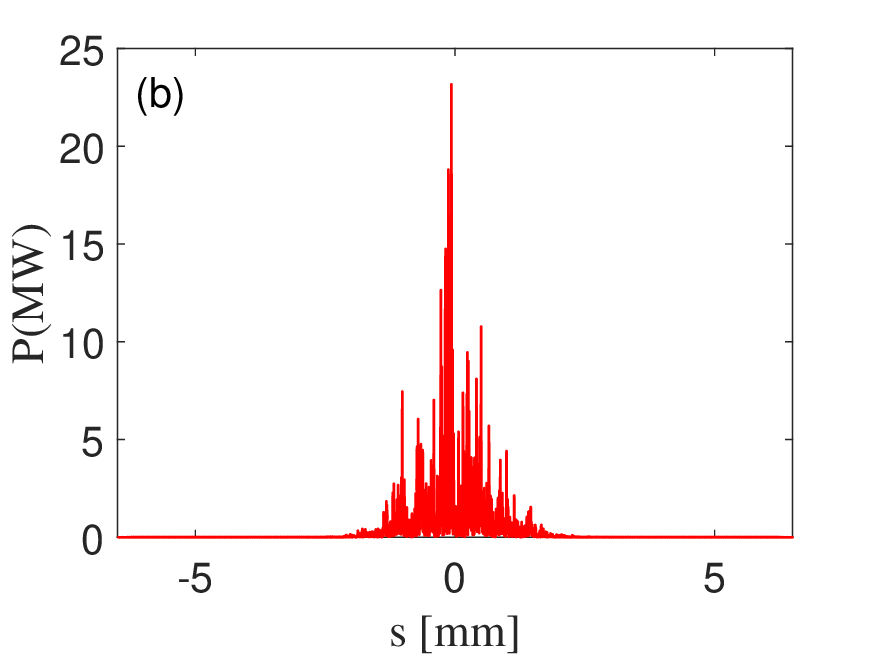}% Here is how to import EPS art
 	\caption{\label{fig:imag4}(a) Gain curve of pulse energy versus the distance  (b) Power within a single pulse }
 \end{figure}

As shown in Fig.\;\ref{fig:imag2}, when the lasing energy loss per electron per turn is controlled below 20 keV, the ratio $ \xi $ increases by no more than 0.1\%, which has a negligible effect on the lasing performance. To be specific, we choose $U_{f}=6.7 $ keV, and thus the new equilibrium energy spread is ($7.3209\times10^{-4}$). The lasing process with this new energy spread value, together with other parameters listed in Table \ref{tab:table3}, is modeled by GENESIS \cite{Reiche1999}. The pulse energy gain curve of a single pulse is shown in Fig.\;\ref{fig:imag4}(a).
 
\begin{figure}[h]
    \includegraphics[scale=0.28]{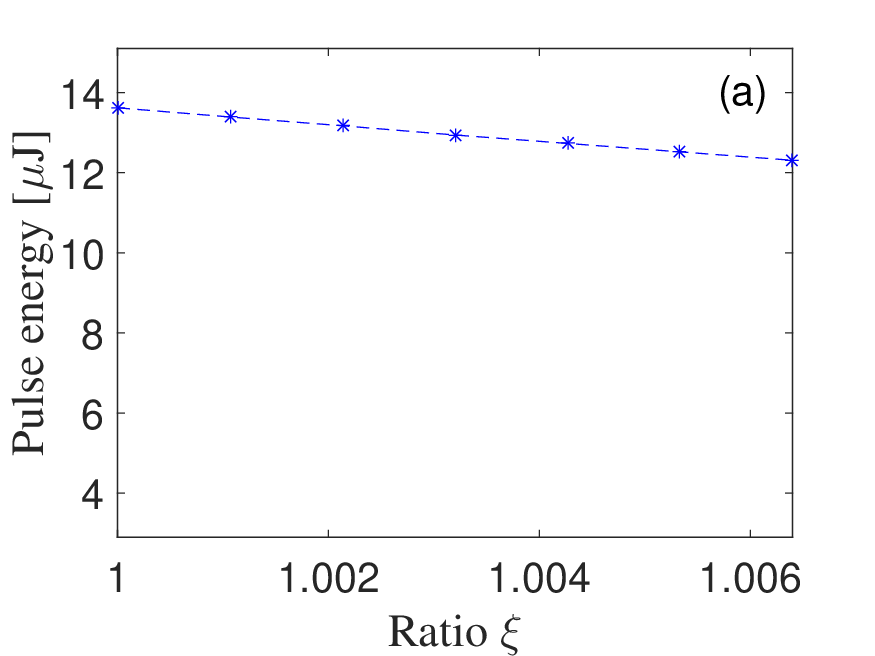}
	\includegraphics[scale=0.28]{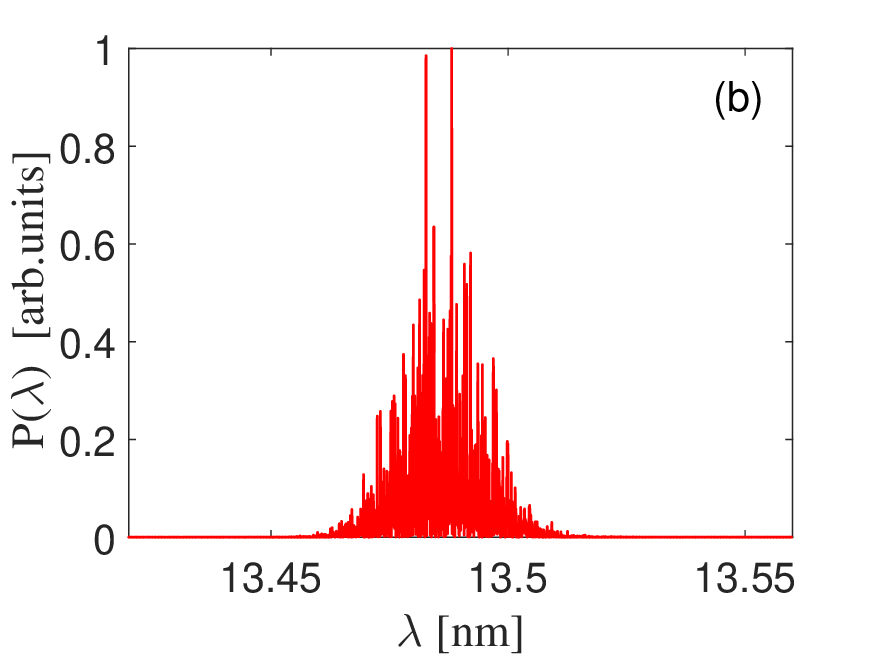}% Here is how to import EPS art
	\caption{\label{fig:imag5} (a) Pulse energy versus Ratio (b) Spectrum of a single pulse }
\end{figure}

The key to establishing sustainable early-stage SASE lasing in the storage ring is to cut off the lasing process at a relatively low power level before the electron beam quality degrades, and hence the electron beam is maintained in an equilibrium state that supports turn-by-turn lasing. In our design, we choose an undulator length of 45.6 m, which is roughly only 13 gain lengths. As illustrated by the shaded region in Fig.\;\ref{fig:imag4}(a), the lasing pulse energy would continue to grow if the undulator extends further, but in the actual design, the undulator terminates at 45.6 m. We could extend the undulator's length for a higher radiation power, at a cost of slightly increased equilibrium beam energy spread as predicted by Eq.\ref{key30}. We verified in simulation that the lasing performance is relatively insensitive to a small change in the beam energy spread, as shown in Fig.\ref{fig:imag5}(a). It is also verified in the GENESIS results that the beam emittance and current have negligible changes in a single pass through the undulator. Overall, the beam parameters are essentially dominated by the design and dynamics of the storage ring as long as the criteria from Eq.(\ref{key30}) are fulfilled, i.e., the radiation losses of the bending magnets and undulators are much larger than the lasing power.  

The temporal distribution of the radiation power at the undulator exit $z=45.6$ m is shown in Fig.\;\ref{fig:imag4}(b). Owing to the bunch length stored in the storage ring being tens of picoseconds which is longer than those in linacs, the pulse energy can be more stable because the longer the electron beam, the more lasing modes within the bunch according to SASE characteristics. The single pulse energy is approximately 13.4 $\upmu$J by integrating over the spectrum. The electron beam is bunched by a 500 MHz RF cavity in the storage ring with an 80\% bucket-filling ratio. Therefore, the average power of the early-stage SASE from each undulator is 5.4 kW (13.4 $\upmu$J $\times500$ MHz $\times0.8$) in the example. The radiation spectrum, as shown in Fig.\ref{fig:imag5}(b), is centered at 13.49 nm with a bandwidth of 0.1\%, which is much narrower than the typical 1-2\% reflectivity bandwidth of multilayer mirrors used for EUVL. 

It is worth mentioning that the spatial coherence of the early-stage SASE is far from that at FEL saturation and close to that of undulator radiation, and hence the techniques developed starting from the mid-1980s for employing synchrotron radiation for lithography are applicable \cite{Becker86, Ehrfeld91}. \\

%\section{Conclusion} 

In summary, in this paper, we propose a new type of SRFEL scheme for generating high average power without seed lasers or optical cavities that can work in a wide spectral range from infrared to EUV. An analytical model is derived to describe this principle. We show an example of this scheme with detailed numerical modeling of the storage ring design and the lasing process for generating multi-kW EUV radiation. The proposed scheme could be a promising tool for future basic study and the manufacturing industry. Further development of this concept and optimization at shorter radiation wavelength towards high average power is an exciting research direction in accelerator physics.

\begin{acknowledgments}
The work was partially supported by the National Key Research and Development Program of China No. 2022YFA1603400.

\end{acknowledgments}

\nocite{*}

%\bibliography{manuscript_kaishang_2023}
\end{document}